\documentclass{article}
\usepackage{amssymb}
\usepackage{fleqn}
\usepackage{epsfig}
\usepackage{graphicx}
\usepackage{amsmath}
\def\beq{\begin{equation}}
\def\eeq{\end{equation}}
\def\bea{\begin{eqnarray}}
\def\eea{\end{eqnarray}}
\def\bq{\begin{quote}}
\def\eq{\end{quote}}

\def\gappeq{\mathrel{\rlap {\raise.5ex\hbox{$>$}}
{\lower.5ex\hbox{$\sim$}}}}
\def\lappeq{\mathrel{\rlap{\raise.5ex\hbox{$<$}}
{\lower.5ex\hbox{$\sim$}}}}
\def\Toprel#1\over#2{\mathrel{\mathop{#2}\limits^{#1}}}

\begin{document}

\pagestyle{empty}
\begin{flushright}
ROME1/1409/05~

DSFNA1/27/2005
\end{flushright}
\vspace*{15mm}

\begin{center}
\textbf{THRESHOLD RESUMMED SPECTRA IN $B\rightarrow X_u l \nu$ DECAYS IN NLO (II)} \\[0pt]

\vspace*{1cm}

\textbf{Ugo Aglietti}\footnote{e-mail address: Ugo.Aglietti@roma1.infn.it} \\[0pt]

\vspace{0.3cm}
Dipartimento di Fisica,\\
Universit\`a di Roma ``La Sapienza'', \\
and I.N.F.N.,
Sezione di Roma, Italy. \\[1pt]

\vspace{0.3cm}
\textbf{Giulia Ricciardi}\footnote{e-mail address:
Giulia.Ricciardi@na.infn.it} \\ [0pt]

\vspace{0.3cm} Dipartimento di Scienze Fisiche,\\
Universit\`a di Napoli ``Federico II'' \\
and I.N.F.N.,
Sezione di Napoli, Italy. \\ [1pt]

\vspace{0.3cm}
\textbf{Giancarlo Ferrera}\footnote{e-mail address: Giancarlo.Ferrera@roma1.infn.it} \\[0pt]

\vspace{0.3cm}
Dipartimento di Fisica,\\
Universit\`a di Roma ``La Sapienza'', \\
and I.N.F.N.,
Sezione di Roma, Italy. \\[1pt]

\vspace*{1cm} \textbf{Abstract} \\[0pt]
\end{center}
We resum to next-to-leading order the distribution in the ratio of the invariant
hadron mass $m_X$ to the total hadron energy $E_X$ and the distribution
in $m_X$ in the semileptonic decays $B\rightarrow X_u l \nu$.
By expanding our formulas, we obtain the coefficients of all the infrared
logarithms at $O(\alpha_S^2)$ and of the leading ones at $O(\alpha_S^3)$.
We explicitly show that the relation between these semileptonic spectra and the
photon spectrum in the radiative decay $B\rightarrow X_s \gamma$ is not
a purely short-distance one.
There are long-distance effects in the semileptonic spectra which are not completely
factorized by the structure function as measured in the radiative decay
and have to be modelled in some way.

\vspace*{4cm} \noindent 

\vfill\eject

\setcounter{page}{1} \pagestyle{plain}

\section{Introduction}

Semileptonic $B$ decays
\begin{equation}
\label{firsteq}
B \, \rightarrow \, X_u \,+ \,  l \, + \, \nu,
\end{equation}
where $X_u$ is any hadron final state coming from the fragmentation of the
$up$ quark, are interesting processes for the study of strong interactions
as well as of weak interactions.
The computation of the spectra in (\ref{firsteq}) is often non trivial because of the
presence of double infrared logarithms in the perturbative expansion, which
formally diverge in the endpoints and therefore must be resummed to all orders
in the QCD coupling $\alpha_S$ \cite{altetal}.
In all generality, large logarithms come from the so-called threshold
region, defined as the one having
\begin{equation}
\label{mXllEX}
m_X \, \ll \, E_X \, \le \, m_b,
\end{equation}
where $m_X$ and $E_X$ are the invariant mass and total energy of the final
hadron state $X_u$ and $m_b$ is the beauty mass.
We find useful to summarize here the main results of \cite{me,noi}.
The infrared logarithms in process (\ref{firsteq})
can be organized in a series of the form
\begin{equation}
\label{start}
\Sigma[u;\,\alpha(Q)] \, = \, 1 \, + \, \sum_{n=1}^\infty
\sum_{k=1}^{2n}
\Sigma_{n k} \, \alpha^n(Q) \, \log^k\left( \frac{1}{u} \right),
\end{equation}
and can be factorized into the universal QCD form factor $\Sigma$.
The $\Sigma_{nk}$'s are numerical coefficients whose explicit expressions
can be obtained from \cite{noi} and $\alpha=\alpha_S$ is the QCD coupling.
$Q$ is the hard scale of the process and is determined by the final hadron
energy:
\begin{equation}
Q \, = \, 2E_X.
\end{equation}
We have defined the hadron variable
\begin{equation}
u \, = \, \frac{1 - \sqrt{1 - \left(2m_X/Q\right)^2} }{1 + \sqrt{1 - \left(2m_X/Q\right)^2} }
\, \simeq \, \left(\frac{m_X}{Q}\right)^2
~~~~~~~~~~~~~~~~~~(0\le u \le 1),
\end{equation}
involving the ratio of the invariant hadron mass to the hard
scale, where in the last member we have taken the leading term in
the threshold region (\ref{mXllEX}) only. As it is well known, the
QCD form factor in  eq.~(\ref{start}) contains at most two
logarithms for each power of $\alpha$, coming from the overlap of
the soft and the collinear region in each emission. Note that the
hard scale $Q$ enters the argument of the infrared logarithms $1/u
\simeq Q^2/m_X^2$ as well as the argument of the running coupling
$\alpha=\alpha(Q)$. One can obtain a factorized form for the
triple differential distribution
--- which is the most general distribution in process (\ref{firsteq}) --- from which
all other spectra can be obtained by integration:
\begin{equation}
\label{tripla}
\frac{1}{\Gamma} \int_0^u \frac{d^3\Gamma}{dx dw du'} \, du'
\,=\, C\left[x,w;\alpha(w\,m_b)\right]\,
\Sigma\left[u;\alpha(w\,m_b)\right] \, + \, D\left[x,u,w;\alpha(w\,m_b)\right],
\end{equation}
where
\begin{equation}
w \,=\, \frac{2 E_X}{m_b}~~~~~~~~(0\le w \le 2);~~~~~~~~~~~~~~~~~~~~~
x \,=\, \frac{2 E_l}{m_b}~~~~~~~(0\le x \le 1).
\end{equation}
$\Gamma$ is the total semileptonic width,
$C\left[x,w;\alpha\right]$ is a short-distance coefficient function
independent on $u$ and
$D\left[x,u,w;\alpha\right]$ is a remainder function not containing
infrared logarithms (i.e. short-distance dominated) and
vanishing for $u\rightarrow 0$ as well as for $\alpha\rightarrow 0$.
The explicit expressions of these functions have been given in \cite{noi}.

The properties of semileptonic decay spectra are best understood
comparing them with the simpler radiative decay
\begin{equation}
\label{raddec}
B \, \rightarrow \, X_s \, + \, \gamma.
\end{equation}
In such decay we have indeed:
\begin{equation}
~~~~~~~~~~~~~~~~~
Q \, = \, m_b \left( 1 - \frac{q^2}{m_b^2} + \frac{m_X^2}{m_b^2} \right)
\, = \,   m_b \left( 1 + \frac{m_X^2}{m_b^2} \right)
\, \simeq \, m_b,
\end{equation}
where $q^{\mu}$ is the 4-momentum of the real photon --- in
general of the probe. In the radiative decays (\ref{raddec}) the
hard scale is therefore independent on the kinematics and is fixed
by the beauty mass. In the semileptonic decay (\ref{firsteq}),
$q^{\mu}$ is the dilepton momentum and we have the more general
situation $0\le q^2 \le m_b^2$; the hard scale is given in this
case by
\begin{equation}
Q \, \simeq \, m_b \left( 1 - \frac{q^2}{m_b^2} \right)
\end{equation}
and depends on the dilepton invariant mass squared $q^2$:
it cannot be identified with the heavy flavor mass $m_b$. Kinematic configurations with
\begin{equation}
m_X \, \ll \, Q \, \approx \, m_b
\end{equation}
as well as with
\begin{equation}
\label{specificSL}
m_X \, \ll \, Q \, \ll \, m_b
\end{equation}
are possible. In fact, in the radiative decays, for example, the
average hadron energy is
\begin{equation}
\langle E_X  \rangle_{rd} \, = \, \frac{1}{2} \, m_b \, \left[ 1 \, + \, O(\alpha) \right],
\end{equation}
while in the semileptonic ones we have the smaller value \cite{noi}
\begin{equation}
\langle E_X  \rangle_{sl} \, = \, \frac{7}{20} \, m_b \, \left[ 1 \, + \, O(\alpha) \right] .
\end{equation}
In general, semileptonic spectra may or may not involve integration over the
hadron energy $E_X$ and, according to \cite{me,noi},
have a different infrared structure in the two cases.
In \cite{noi} we have studied in detail the simpler case of the distributions
not integrated over the hadron energy, i.e. over the hard scale $Q$,
which have the same infrared structure
of the hadron invariant mass distribution in the radiative decay (\ref{raddec}):
\begin{equation}
\label{specraddec}
\frac{1}{\Gamma_R}\int_0^{t_s}\frac{d\Gamma_R}{dt_s'} \, dt_s'
\, = \,
C_R(\alpha) \left( 1 \, + \,
\, \sum_{n=1}^\infty
\sum_{k=1}^{2n}
\Sigma_{n k} \, \alpha(m_b)^n\, \log^k \frac{1}{t_s} \right)
\, + \, D_R(t_s;\alpha).
\end{equation}
We have defined $t_s = m_{X_s}^2/m_b^2 = 1-x_{\gamma}$,
with $x_{\gamma}=2E_{\gamma}/m_b$,
$\Gamma_R$ is the total radiative width, $C_R(\alpha)$ is a short-distance
coefficient function and $D_R(t_s;\alpha)$ is a short-distance remainder function.

In this paper, we attack the distributions integrated over the hadron energy,
which have a more complicated infrared structure than the one in (\ref{raddec}).

In sec.~\ref{sect1} we resum to next-to-leading order (NLO) the distribution
in the hadron variable $u$ defined before.
The infrared logarithms appearing in the perturbative expansion of this spectrum,
\begin{equation}
\frac{1}{\Gamma}\int_0^u\frac{d\Gamma}{du'} \, du' \, = \, C_U(\alpha)\,
\left( 1 \, + \, \sum_{n=1}^\infty
\sum_{k=1}^{2n}
\Sigma_{U nk} \, \alpha\left(m_b\right)^n \, \log^k \frac{1}{u} \right)
\, + \, D_U(u;\alpha),
\end{equation}
coincide in first order with those in the decay (\ref{raddec}):
$\Sigma_{U12}=\Sigma_{12}$ and $\Sigma_{U11}=\Sigma_{11}$,
while they differ in higher orders.
$C_U(\alpha)$ is a short-distance coefficient function and $D_U(u;\alpha)$
is a short-distance remainder function whose explicit expressions will
be given in sec.~(\ref{sect1}).

In sec.~\ref{sect2} we compute to NLO the distribution in the variable
\begin{equation}
t \, = \, \frac{m_X^2}{m_b^2}
~~~~~~~~~~~~~~~~~~~~~~~~~~(0\le t \le 1),
\end{equation}
i.e. the distribution in the invariant hadron mass squared.
The infrared logarithms appearing in this distribution,
\begin{equation}
\frac{1}{\Gamma}\int_0^t\frac{d\Gamma}{dt'} \, dt' \, = C_T(\alpha)\,
\left( 1 \, + \, \sum_{n=1}^\infty
\sum_{k=1}^{2n}
\Sigma_{T nk} \, \alpha\left(m_b\right)^n \, \log^k \frac{1}{t}  \right)
\, + \, D_T(t;\alpha),
\end{equation}
differ also at the $O(\alpha)$ single logarithm level from the
corresponding ones in (\ref{raddec}):  $\Sigma_{T11}\ne\Sigma_{11}$.
We also define a non-minimal factorization-resummation scheme
which seems to have better convergence properties of the perturbative
series than the minimal one.

We define for both spectra effective form factors which resum
the large logarithmic corrections.
These form factors involve a convolution
of a process-dependent coefficient function with the universal
QCD form factor $\Sigma$ entering the radiative decay (\ref{raddec}) and the
triple differential distribution in the decay (\ref{firsteq}).
The definition of the effective form factors we give is, in a sense,
non perturbative, because we do not look at the explicit perturbative
expansion of $\Sigma$.

Finally, in sec.~\ref{concl}, we draw our conclusions and we consider
generalizations of our results.

\section{Distribution in the hadron mass/energy ratio}
\label{sect1}

In this section we compute the resummed distribution in the variable $u$
(defined in the introduction) to next-to-leading order (NLO).
That is accomplished by integrating the resummed double distribution
in $u$ and $w$ obtained in sec.~(4) of \cite{noi} over $w$:
\begin{equation}
\label{firstu}
\frac{1}{\Gamma}\frac{d\Gamma}{du}
\, = \, \int_0^{1+u} dw \, \frac{1}{\Gamma}\frac{d^2\Gamma}{dw du}.
\end{equation}
By replacing the resummed expression on the r.h.s. of eq.~(\ref{firstu}),
we obtain:
\begin{equation}
\label{exactU}
\frac{1}{\Gamma}\frac{d\Gamma}{du}
\, = \, \int_0^{1+u} dw \, C_H(w;\,\alpha) \, \sigma\left[u;\,\alpha(w\,m_b)\right]
\, + \, \int_0^{1+u} dw \, d_H(u,w;\,\alpha).
\end{equation}
$C_H(w;\,\alpha)$ is a short-distance coefficient function having an
expansion in powers of $\alpha$:
\begin{equation}
\label{CHw}
C_H(w;\,\alpha) \, = \, C_H^{(0)}(w) \, + \, \alpha \, C_H^{(1)}(w)
\, + \, \alpha^2 \, C_H^{(2)}(w) \, + \, O(\alpha^3),
\end{equation}
with
\begin{eqnarray}
\label{CHw1}
C_H^{(0)}(w) &=& 2w^2(3-2w);
\\
\label{CHw2}
C_H^{(1)}(w) &=& \frac{C_F}{\pi} \, 2w^2(3-2w)
\left[ {\rm Li}_2(w) \, + \,\log w\log(1-w) \, - \, \frac{35}{8}
\, - \, \frac{9-4w}{6-4w} \log w \right],
\end{eqnarray}
where $C_F=(N_c^2-1)/(2N_c)$, $N_c=3$ is the number of colors and
$\alpha=\alpha(m_b)$. $\sigma\left(u;\,\alpha\right)$ is the
differential QCD form factor:
\begin{equation}
\sigma(u;\,\alpha) \, = \, \frac{d}{du} \Sigma(u;\,\alpha),
\end{equation}
with $\Sigma(u;\,\alpha)$ being the cumulative form factor
considered in the introduction,
\begin{equation}
\label{Sigma}
\Sigma(u;\,\alpha) \, = \, 1 \, + \, \alpha \, \Sigma^{(1)}(u)
\, + \, \alpha^2 \, \Sigma^{(2)}(u) \, + \, O(\alpha^3)
\end{equation}
and
\begin{equation}
\Sigma^{(1)}(u)\, = \, - \, \frac{C_F}{\pi}
\left[ \frac{1}{2}\log^2 u \, + \, \frac{7}{4}\log u \right].
\end{equation}
Finally, $d_H(u,w;\,\alpha)$ is a short-distance remainder function whose explicit
expression is not needed here.
We are interested in the threshold region (\ref{mXllEX}), which can also be defined
as the one having
\begin{equation}
u \, \ll \, 1.
\end{equation}
Since large logarithms originate only from the first integral on the r.h.s.
of eq.~(\ref{exactU}),
in order to isolate them, let us neglect at first the contribution from the
remainder function $d_H(u,w;\,\alpha)$.
The small terms for $u\rightarrow 0$ will be included later on, by expanding
the resummed expression and comparing with fixed-order spectrum,
as discussed at the end of sec.~2 of \cite{noi}.
The integration of the first term on the r.h.s. of eq.~(\ref{exactU}) over the hadron energy
is not trivial because the coefficient function $C_H(w;\,\alpha)$ depends on $w$
as well as the QCD form factor $\sigma[u;\,\alpha(w\,m_b)]$, which depends on
$w$ via the scale of the coupling $\alpha=\alpha(w\,m_b)$.
Unlike the distributions considered in \cite{noi}, we cannot factor out in this case
the universal form factor $\sigma$.

Since it is technically simpler to deal with partially-integrated form factors
rather than with differential ones, let us define the event fraction
\begin{equation}
R_U(u) \, = \, \int_0^u  \frac{1}{\Gamma}\frac{d\Gamma}{du'} du',
\end{equation}
which has the end-point values:
\begin{equation}
R_U(0) \, = \,0;~~~~~~~~~~R_U(1) \, = \,1.
\end{equation}
The spectrum in $u$ is trivially obtained by differentiation:
\begin{equation}
\frac{1}{\Gamma}\frac{d\Gamma}{du} \, = \, \frac{d}{du} \, R_U(u).
\end{equation}
Integrating both sides of eq.~(\ref{exactU}), we obtain:
\begin{eqnarray}
R_U\left[u;\alpha(m_b)\right]
&=& \int_0^u du' \int_0^{1+u'} \, dw \, C_H(w;\alpha)\,\sigma[u';\alpha(w\,m_b)]
\, + \, O(u,\alpha)
\nonumber\\
&=& \int_0^u du' \int_0^1 \, dw \, C_H(w;\alpha)\,\sigma[u';\alpha(w\,m_b)]
 + \int_0^u du' \int_1^{1+u'} \, dw \, C_H(w;\alpha)\,\sigma[u';\alpha(w\,m_b)]+
\nonumber\\
& & + \, O(u;\alpha),
\end{eqnarray}
where by $O(u;\alpha)$ we denote terms which vanish for
$u\rightarrow 0$ as well as for $\alpha\rightarrow 0$. The second
integral in the last member of the r.h.s. extends to a kinematic
region in $w$ which is $O(u)$ and therefore can be dropped in the
threshold region; we can therefore assume tree-level kinematics:
\begin{equation}
0 \, \le \, w \, \le \, 1.
\end{equation}
By exchanging the order of the integrations in the remaining integral, we obtain:
\begin{equation}
\label{Unorem}
R_U\left[u;\alpha(m_b)\right]\,=\,\int_0^1 \, dw \, C_H(w;\alpha)\,\Sigma[u;\alpha(w\,m_b)]
\, + \, O(u;\alpha).
\end{equation}
Substituting the explicit expressions for the coefficient function given in eq.~(\ref{CHw})
and of the QCD form factor given in eq.~(\ref{Sigma}), we obtain:
\begin{equation}
\int_0^1 \, dw \, C_H(w;\alpha)\,\Sigma(u;\,\alpha) \, = \,
1 \, - \,
\frac{\alpha\,C_F}{\pi}
\left[
\frac{1}{2}\log^2 u \, + \, \frac{7}{4}\log u \, + \, \frac{335}{144}
\right]
\, + \, O(\alpha^2).
\end{equation}
The next step is to factorize the event fraction $R_U(u;\,\alpha)$
into:
\begin{itemize}
\item
a QCD form factor $\Sigma_U(u;\,\alpha)$ containing the long-distance contributions,
i.e. the $\log 1/u$ terms diverging for $u\rightarrow 0$;
\item
a coefficient function $C_U(\alpha)$ containing the constant
terms for $u\rightarrow 0$;
\item
a remainder function $D_U(u;\,\alpha)$, collecting the left-over small
contributions $O(u;\alpha)$, vanishing for $u\rightarrow 0$ and for $\alpha\rightarrow 0$.
\end{itemize}
Let us write therefore:
\begin{equation}
\label{resumU} R_U\left(u;\,\alpha\right) \, = \, C_U(\alpha) \,
\Sigma_U(u;\,\alpha) \, + \, D_U(u,\,\alpha).
\end{equation}
The coefficient function, the form factor and the remainder function can
all be expanded in powers of $\alpha$:
\begin{eqnarray}
C_U(\alpha) &=& 1 \, + \, \alpha \, C_U^{(1)} \, + \, \alpha^2 \, C_U^{(2)}
\,+ \, O(\alpha^3);
\\
\Sigma_U(u;\,\alpha) &=& 1 \, + \, \alpha \, \Sigma_U^{(1)}(u) \, + \alpha^2 \, \Sigma_U^{(2)}(u)
\, + \, O(\alpha^3);
\\
D_U(u;\,\alpha) &=& \alpha \, D_U^{(1)}(u) \, + \alpha^2 \, D_U^{(2)}(u)
\, + \, O(\alpha^3).
\end{eqnarray}
The knowledge of soft-gluon dynamics allows the resummation of the
dominant terms to all orders in $\alpha$ in $\Sigma_U$; this is
instead not possible for the coefficient function and the
remainder function, which are not long-distance dominated and for
them one has to use truncated expansions.

The above conditions do not completely specify the form of the
form factor, of the coefficient function and of the remainder
function, so we have to select a factorization scheme. Let us
choose a minimal scheme, in which the form factor contains {\it
only} logarithmic terms:
\begin{equation}
\Sigma_U(u;\,\alpha) \, = \, 1 \, + \, \sum_{n=1}^{\infty}\sum_{k=1}^{2n}
\Sigma_{Unk} \, \alpha^n \, L^k.
\end{equation}
where
\begin{equation}
L \, \equiv \, \log \frac{1}{u}.
\end{equation}
From the above definition, it follows that $\Sigma_U$ has the same normalization
as $\Sigma$:
\begin{equation}
\label{SigmaU1}
\Sigma_U(1;\,\alpha)\,=\,1,
\end{equation}
which holds to any order in $\alpha$.
We obtain in first order:
\begin{eqnarray}
\Sigma_U^{(1)}(u) &=&
- \, \frac{C_F}{\pi}
\left[
\frac{1}{2}\log^2 u \, + \, \frac{7}{4}\log u
\right];
\\
\label{CU1}
C_U^{(1)}  &=& - \, \frac{335}{144} \frac{C_F}{\pi}.
\end{eqnarray}
Note that
\begin{equation}
\label{onlyfirst}
\Sigma_U^{(1)}(u)=\Sigma^{(1)}(u).
\end{equation}
As we are going to explicitly show later in this section, this
property does not hold in higher orders of $\alpha$. For
$\alpha(m_b)=0.22$, we have a first order correction to the
coefficient function of $-\,21.7\%$.

The remainder function is obtained by imposing consistency
between the resummed expression and the fixed-order one (matching).
By expanding to first order in $\alpha$ the resummed expression
(\ref{resumU})
and imposing the equality with the full $O(\alpha)$ result,
one obtains \cite{ucg}:
\begin{equation}
\label{restoU}
D_U^{(1)}(u) \, = \, \frac{C_F}{\pi}
\Big[
\frac{ u(15624 - 2688 u - 1352 u^2 + 141 u^3) }{5040}
- \frac{21-84 u - 29 u^2 + 6 u^3}{210} u \log u
\Big].
\end{equation}
As required, the remainder function goes to zero in the elastic point $u=0$
(as $u\log u$).
Taking $u=1$ in eq.~(\ref{resumU}), we obtain the following relation
between the coefficient function and the remainder function in the upper
endpoint:
\begin{equation}
C_U(\alpha) \, = \, 1 \, - \, D_U(1;\alpha),
\end{equation}
which holds to any order in $\alpha$ and is verified in first order
(see eqs.~(\ref{CU1}) and (\ref{restoU})).

The minimal scheme has been defined above by looking at the explicit form
of the event fraction $R_U(u;\,\alpha)$ as a power series in $\alpha$:
one reorganizes the series picking up the logarithmic terms and putting them
into the effective form factor $\Sigma_U$. It is also possible to give a different
definition of the minimal scheme which does not make use
of the explicit expansion of the event fraction.
Since
\begin{equation}
\int_0^1 \, dw \, C_H(w;\alpha)\,\Sigma\left[u;\,\alpha(w\,m_b)\right]
\, = \, C_U(\alpha) \, \Sigma_U(u;\,\alpha)
\end{equation}
and
\begin{equation}
\Sigma_U(1;\,\alpha) \, = \, 1,
\end{equation}
we can define the coefficient function to all orders as:
\begin{equation}
\label{defCU}
C_U(\alpha) \, = \, \int_0^1 dw \, C_H(w;\,\alpha).
\end{equation}
The effective form factor is therefore written as:
\begin{equation}
\label{defSigU}
\Sigma_U\left(u;\,\alpha\right) \,=\,
\frac{\int_0^1 dw \, C_H(w;\,\alpha) \, \Sigma[u;\,\alpha(w\,m_b)]}
     { \int_0^1 dw \, C_H(w;\,\alpha) }.
\end{equation}
The definition (\ref{defSigU}) of the minimal scheme has the following phenomenological
advantage. One may wish to use for $\Sigma(u;\alpha)$, instead of the perturbative expression,
for example the result of a fit to some experimental data or a non-perturbative model\footnote{
For the coefficient function, one can still use
the perturbative result. That is because the coefficient function,
unlike the form factor, is short-distance dominated, and therefore its perturbative
evaluation is more reliable.}.
In these cases, $\Sigma(u;w)$ does not depend on the coupling and therefore cannot be
expanded in $\alpha$, but the effective form factor $\Sigma_U$ in the minimal
scheme can still be computed by means of eq.~(\ref{defSigU}).

The representation of the effective form factor $\Sigma_U$ given in
eq.~(\ref{defSigU}) allows us to make a few general comments:
\begin{itemize}
\item
$\Sigma_U(u;\,\alpha)$ factorizes all the
threshold logarithms in the spectrum but is, unlike $\Sigma(u)$,
process dependent. That is because it involves the convolution of the universal
form factor $\Sigma(u;\,\alpha)$ with the process-dependent coefficient function
over all the hadron energies.
$C_H(w;\,\alpha)$ has the role of a probability distribution:
it gives the probability for the hadronic subprocess with hard scale
$Q\,=\,w\,m_b$ to occur.
In the $u$ distribution hadronic subprocesses with all the possible
hard scales $Q$ from zero up to $m_b$ do contribute, while in the radiative decay
(\ref{raddec}) $Q$ is kinematically fixed to the upper value $m_b$.
The relation between these two spectra therefore is not a purely short-distance
one: to relate these two distributions one has to model in some way the variation
of the form factor $\Sigma[u;\alpha(Q)]$ with $Q$ ranging from $m_b$ down to zero.
In agreement with physical intuition, the problem in the computation of the $u$
distribution is the estimate of the contributions from small hard scales,
$Q \ll m_b$, where perturbation theory is expected to fail.
However, since $C_H \, \propto \, Q^2$, small hard scales give a small
contribution to the total.
We may say that the $u$ spectrum is ``protected'' from long-distance
effects related to large logarithms with a large coupling, i.e.
related to the region in eq.~(\ref{specificSL});
\item
since $\alpha(w\,m_b)\,=\,\alpha(m_b)\,+\,O(\alpha^2)$,
the effective form factor and the universal form factor coincide in
first order, as already found by explicit computation (see eq.~(\ref{onlyfirst}));
\item
were not for the dependence of the form factor $\Sigma\,=\,\Sigma[u;\,\alpha(w\,m_b)]$
on $w$ through the running coupling, $\alpha\,=\,\alpha(w\,m_b)$,
the effective form factor $\Sigma_U$ would coincide with the universal one $\Sigma$
to all orders.
\end{itemize}

Let us now discuss the higher orders in the perturbative expansion
of $\Sigma_U(u;\,\alpha)$.
One has to insert in eq.~({\ref{defSigU}}):
\begin{itemize}
\item the QCD form factor $\Sigma$, whose next-to-next-to-leading
order corrections (NNLO) are now well established due to the
recent revaluation of the resummation constant $D_2$ in
\cite{bsgamma3}; \item the coefficient function $C_H(w;\,\alpha)$,
which is known to $O(\alpha)$, i.e. to NLO (cfr. eq.~(\ref{CHw})),
\end{itemize}
and perform the integration over $w$. To obtain the truncated
expansion of $\Sigma_U$, one simply replaces the truncated
expansions of $\Sigma$ and of $C_H$, expands the product and
integrates term by term.

The exponential structure of the threshold logarithms in $\Sigma(u;\,\alpha)$
is partially spoiled in $\Sigma_U(u;\,\alpha)$
because of the integration over the hard scale $Q\, = \, w \, m_b$,
but it is not completely ruined.
In order to simplify the representation of the large logarithms, it is
therefore convenient to introduce the
exponent of the form factor also in the effective case,
i.e. to define $G_U$ as:
\begin{equation}
\Sigma_U \, = \, e^{ G_U }.
\end{equation}
Expanding the exponent in powers of $\alpha$ up to third order
included, we have an expansion of the same form of $G$, defined in
\cite{noi}, that is:
\begin{equation}
G_U\left(u;\,\alpha\right) \,=\, \sum_{n=1}^{\infty} \sum_{k=1}^{n+1}
G_{U n k}\, \alpha^n \, L^k
\,=\,  G_{U 12} \alpha L^2 + G_{U 11} \alpha L
+ G_{U 23} \alpha^2 L^3  + G_{U 22} \alpha^2 L^2
+ G_{U 21} \alpha^2 L + \cdots,
\end{equation}
where
\begin{equation}
L \,  \equiv \, \log\frac{1}{u} \, \ge \, 0
\end{equation}
and $\alpha \,\equiv\, \alpha(m_b)$.
In order to explicitly see the effects of the integration over the
hard scale, let us give the coefficients $G_{Uij}$ up to
the third order in terms of the corresponding
coefficients $G_{ij}$ of the radiative decay (\ref{raddec}) and of the
resummation constants $A_i,~B_i,~D_i$ and of the $\beta$-function
coefficients $\beta_i$ \cite{noi}:
\begin{eqnarray}
G_{U 12} &=& G_{12};
\\
G_{U 11} &=& G_{11};
\\
G_{U 23} &=& G_{23};
\\
G_{U 22} &=& G_{22} - \frac{5}{12} A_1\beta_0;
\\
G_{U 21} &=& G_{21} - \frac{5}{6} \beta_0 \Big( B_1 + D_1 \Big);
\\
G_{U 34} &=& G_{34};
\\
G_{U 33} &=& G_{33} - \frac{5}{6} A_1 \beta_0^2;
\\
G_{U 32} &=& G_{32}
- \frac{5}{6}\,A_2\,\beta_0
- \frac{1}{36}\,A_1\,
   \Big(
      23\,{\beta_0}^2 + 15\,\beta_1
     \Big)
-  \frac{5}{6}{\beta_0}^2\,
   \Big( B_1 + 2\,D_1 \Big)  -
  \frac{5}{6}  \,{A_1}^2\,
     \beta_0\,z(2)   +
\nonumber\\
&-&
        \left(
\frac{547}{216} - 2\,z(3)
       \right)
\frac{C_F}{\pi } \, A_1\,
     \beta_0\,,
\end{eqnarray}
where $z(a)=\sum_{n=1}^{\infty}1/n^a$ is the Riemann Zeta function
and the explicit expressions of the $G_{ij}$ have been given in
sec.~(2) of \cite{noi}. For a reference use in phenomenological
analysis, let us also give the explicit expressions for the
coefficients, which can be checked against the second and third
order computations of the $u$ spectrum as soon as the latter will
become available. We report only the coefficients differing from
the starting ones, $G_{U i j} \ne G_{i j}$:
\begin{eqnarray}
G_{U 22} &=&\frac{C_F}{{\pi }^2} \,
    \Bigg[ - \frac{n_f}
       {48} - \frac{C_F\,z(2)}{2} +
      C_A\,
       \left( - \frac{5}{96}  +
         \frac{z(2)}{4} \right)
      \Bigg]  ;
\\
G_{U 21} &=&
\frac{C_F }{\pi^2} \,
\Bigg[ n_f\,
       \left( - \frac{5}{6} +
         \frac{z(2)}{6} \right)
       + C_A\,
       \left( \frac{215}{48} -
         \frac{17\,z(2)}{12} -
         \frac{z(3)}{4}
         \right)  +
      C_F\,
       \left( \frac{3}{32} +
         z(2) +
         \frac{z(3)}{2}
         \right)  \Bigg]  ;
\\
G_{U 33} &=&
\frac{C_F }{\pi^3} \,
\Bigg[
-\frac{11}{432}\,n_f^2 + C_A\,n_f \left(\frac{95}{216}
      -\frac{z(2)}{12}\right)+
    C_A^2\left(-\frac{2471}{1728} + \frac{11\,z(2)}{24}\right) +\nonumber\\
 &+&   C_F\,n_f \left(\frac{1}{16}+\frac{z(2)}{4}\right)
  - \frac{11}{8}\,C_F\,C_A\,z(2) +
    \frac{1}{3}\,C_F^2\,z(3)
 \Bigg]  ;
\\
G_{U 32} &=&
\frac{C_F }{\pi^3} \,
\Bigg[
\,n_f^2 \left( \frac{49}{432}  - \frac{z(2)}{36} \right) +
    C_F\,C_A \left(-\frac{23177}{10368} + \frac{155\,z(2)}{96}
     - \frac{11\,z(3)}{12}+ \frac{5\,z(4)}{4} \right) + \nonumber\\
&+& C_F\,n_f \left(\frac{2971}{5184}-\frac{17\,z(2)}{48}
   - \frac{z(3)}{12}\right) +
    C_F^2\left( - \frac{7\,z(3)}{4}+ \frac{z(4)}{4} \right) +
  C_A\,n_f\left(-\frac{1709}{1728} + \frac{19 z(2)}{72}-\frac{z(3)}{24}\right) +
\nonumber \\
&+&    C_A^2\left(\frac{1541}{864} - \frac{4\,z(2)}{9}
     + \frac{77\,z(3)}{48} -\frac{11\,z(4)}{16} \right)
 \Bigg],
\end{eqnarray}
where $C_A=N_c=3$ and $n_f$ is the number of active flavors.
Let us make a few comments about the results obtained:
\begin{itemize}
\item it is remarkable that we could explicitly compute $G_{U 32}$
with the knowledge of the first two orders of the coefficient
function $C_H(w;\,\alpha)$ only (cfr. eqs.~(\ref{CHw1}) and
(\ref{CHw2})). We can compute also the coefficients $G_{U n,n-1}$
for $n>3$, i.e. the exponent $G_U$ to NNLO. As explained in detail
in \cite{cattren1, noi}, NNLO means indeed that for each power of
$\alpha$ we can compute the three principal logarithms:\footnote{
If we look instead at the form factor itself, $\Sigma_U$, NNLO
means that we can compute for any $n$ all the coefficients
$\Sigma_{Unk}$ with $n-1 \le k\le 2n$:
\begin{equation}
\Sigma_{U n,2n}\,\alpha^n \, L^{2n},~~~~~
\Sigma_{U n,2n-1}\,\alpha^n \, L^{2n-1},~~~~~
\cdots,~~~~~
\Sigma_{U n,n}\,\alpha^n \, L^n,~~~~~
\Sigma_{U n,n-1}\,\alpha^n \, L^{n-1}.~~~
\end{equation}
For instance for $n=3$ we can compute the coefficients of the logarithms
from power six down to power two included.}
\begin{equation}
G_{U n,n+1}\,\alpha^n \, L^{n+1},~~~~~
G_{U n,n}\,\alpha^n \, L^n,~~~~~
G_{U n,n-1}\,\alpha^n L^{n-1}.
\end{equation}
By general counting arguments, one would expect that the NNLO
corrections to $\Sigma_U$ also require the knowledge of the NNLO
contribution to the coefficient function, i.e. of the
$O(\alpha^2)$ term $C_H^{(2)}(w)$. That is actually not the case
because the NNLO contributions proportional to $C_H^{(2)}(w)$
cancel between the numerator and the denominator in the definition
of the effective form factor in eq.~(\ref{defSigU}). However, a
complete NNLO resummation of the $u$-spectrum also requires the
knowledge of the second-order correction to the coefficient
function $C_U^{(2)}$, and for that $C_H^{(2)}(w)$ is needed (see
eq.(\ref{defCU})); \item Since the coefficients of the threshold
logarithms in the $u$ distribution and in the photon spectrum in
the radiative decay (\ref{raddec}) differ from two loops on in
NLO, the cancellation of long-distance effects in the ratio
considered in \cite{ucg}
\begin{equation}
\label{ldonly}
R(u)\, = \, \frac{ d\Gamma_R/dt_s(t_s=u) }{ d\Gamma/du(u) }
\end{equation}
is not exact but occurs only in leading order. As previously
discussed, the $u$ distribution has additional long distance
effects with respect to the radiative decay related to small
hadron energies.
\end{itemize}

The differential spectrum in $u$ is obtained from the event fraction
$R_U(u)$ by differentiation:
\begin{equation}
\frac{1}{\Gamma}\frac{d\Gamma}{du} \, = \, C_U(\alpha)
\, \sigma_U\left(u;\,\alpha\right) \, + \, d_U(u;\,\alpha),
\end{equation}
where we have defined
\begin{equation}
\sigma_U \left(u;\alpha\right)\,\equiv
\,\frac{d}{du} \Sigma_U \left(u;\alpha\right);
~~~~~~~~~~~
d_U(u;\alpha) \, \equiv \, \frac{d}{du} \, D_U(u;\alpha).
\end{equation}
The coefficient function is clearly the same in the partially-integrated
spectrum and in the differential one, while the remainder function is obtained
by differentiation and reads:
\begin{equation}
d_U^{(1)}(u) \, = \, \frac{C_F}{\pi}
\left[
\frac{36-8u-8u^2+u^3}{12} -
\frac{ 7 - 56u - 29u^2 + 8u^3}{70} \log u
\right].
\end{equation}

\section{Hadron mass spectrum}
\label{sect2}

In this section we resum to NLO the distribution
in the invariant hadron mass squared, i.e. the distribution in the
variable
\begin{equation}
\label{deft}
~~~~~~~~~~~~~~~
t \,=\, \frac{m_X^2}{m_b^2} \, = \,
\frac{u \, w^2}{(1+u)^2}\, \simeq \, u\, w^2 ~~~~~~~~~~~~~~~~~~~~~~~ (0\le t\le 1),
\end{equation}
where in the last member we have kept the leading term for
$u\rightarrow 0$ only.
This distribution is obtained by integrating the distribution
in the hadron variables $u$ and $w$ with the previous kinematic
constraint:
\begin{equation}
\label{firstt}
\frac{1}{\Gamma}\frac{d\Gamma}{dt} \, = \,
\int_D du dw  \, \frac{1}{\Gamma}\frac{d^2\Gamma}{du dw}
\delta\left[t-\frac{u \, w^2}{(1+u)^2}\right];
\end{equation}
the integration covers the whole phase space $D$ of the hadron
variables:
\begin{equation}
  \int_D du dw
\,=\, \int_0^2 dw \int_{max[0,w-1]}^1 du
\,=\, \int_0^1 du \int_0^{1+u} dw.
\end{equation}
It is convenient to evaluate the event fraction, defined
like in the previous distribution as:
\begin{equation}
R_T(t)\,=\,\int_0^t dt' \, \frac{1}{\Gamma}\frac{d\Gamma}{dt'}\,,
\end{equation}
with the end-point values $R_T(0)=0$ and $R_T(1)=1$.
Integrating both sides of eq.~(\ref{firstt}), one obtains:
\begin{equation}
R_T(t) \,=\,  \int_D du dw  \, \frac{1}{\Gamma}\frac{d^2\Gamma}{du dw}
\theta \left[t-\frac{u \, w^2}{(1+u)^2}\right].
\end{equation}
By inserting the resummed form for the double hadron distribution
and neglecting at first the remainder function, we obtain:
\begin{equation}
\label{RT}
R_T(t;\,\alpha) \,=\,
\int_D du dw \, C_H(w;\,\alpha) \, \sigma\left[u;\,\alpha(w\,m_b)\right]\,
\theta \left[t-\frac{u \, w^2}{(1+u)^2}\right] \, + \, O(t,\alpha),
\end{equation}
where by $O(t,\alpha)$ we denote terms which are zero at the threshold
and vanish for $\alpha=0$.
In order to isolate the large logarithms, let us simplify the domain $D$
into the unit square
\begin{equation}
0 \, \le \, u, \, w \, \le \,1
\end{equation}
and  simplify the kinematic constraint as well as:
\begin{equation}
\theta \left[t - u \, w^2/(1+u)^2\right]
\,\, \rightarrow \,\,
\theta[t - u \, w^2].
\end{equation}
Let us observe that the variable $t$ keeps unitary range even after
such approximations. We then obtain:
\begin{eqnarray}
R_T(t;\alpha) &=&  \int_0^1 \int_0^1
du dw \, C_H(w;\alpha) \, \sigma\left[u;\alpha(w\,m_b)\right]\,
\theta \left[t- u w^2\right] \, + \, O(t,\alpha)
\nonumber\\
&=&
\label{goodsplit}
\int_0^{\sqrt{t}} dw \, C_H(w;\alpha) \, + \,
\int_{\sqrt{t}}^1 dw \, C_H(w;\alpha) \,
\Sigma\left[t/w^2;\,\alpha(w\,m_b)\right] \, + \, O(t,\alpha).
\end{eqnarray}
Let us remark that the present  case is more involved with respect to the one
treated in the previous section, because the hadron mass squared $t$ is a combination
of both variables used for threshold resummation, namely $u$ and $w$.
Neglecting infinitesimal terms for $t\rightarrow 0$,
the expression above can be further simplified by neglecting the first integral
and integrating the second integrand over $w$ down to zero,
\begin{equation}
R_T(t;\,\alpha) \, = \,
\int_0^1
dw \, C_H(w;\alpha) \, \Sigma\left[t/w^2;\,\alpha(w\,m_b)\right] \, + \, O(\alpha,t).
\end{equation}
By inserting the first-order expressions for the coefficient function and
the form factor given in the previous section, we obtain:
\begin{equation}
\int_0^1
dw \, C_H(w;\alpha) \, \Sigma\left[t/w^2;\,\alpha(w\,m_b)\right]  \, = \,
1\, - \, \frac{\alpha \, C_F}{\pi}
\left[
\frac{1}{2}\log^2 t
\, + \, \frac{31}{12}\log t
\, + \, \frac{637}{144}
\right]
\, + \, O(\alpha^2).
\end{equation}
As anticipated in the introduction, we have a different coefficient for the single
logarithm with respect to the hadron mass distribution in the radiative decay (\ref{raddec})
or the $u$ distribution of the previous section.

As with the $u$ distribution, we introduce a resummed form of the event fraction as:
\begin{equation}
\label{R_T}
R_T(t;\,\alpha) \,=\, C_T(\alpha) \, \Sigma_T(t;\,\alpha) \,
+ \, D_T(t;\,\alpha).
\end{equation}
All the functions above have an expansion in powers of $\alpha$:
\begin{eqnarray}
C_T(\alpha) &=& 1 \, + \, \alpha \, C_T^{(1)} \, + \, \alpha^2 \, C_T^{(2)}
\,+ \, O(\alpha^3);
\\
\Sigma_T(u;\,\alpha) &=& 1 \, + \, \alpha \, \Sigma_T^{(1)}(t) \, + \alpha^2 \, \Sigma_T^{(2)}(t)
\, + \, O(\alpha^3);
\\
D_T(t;\,\alpha) &=& \alpha \, D_T^{(1)}(t) \, + \alpha^2 \, D_T^{(2)}(t)
\, + \, O(\alpha^3).
\end{eqnarray}
Let us consider a minimal factorization scheme, where
only logarithms in $t$ are factorized in
the effective form factor $\Sigma_T$.
Since we will consider a different scheme later on in this section, let
us denote the quantities in the minimal scheme with a bar.
The first-order corrections to the form factor and the coefficient function
in the minimal scheme read:
\begin{eqnarray}
\bar{\Sigma}_T^{(1)}(t) &=&
- \, \frac{C_F}{\pi}
\left[
\, \frac{1}{2}\log^2 t
\, + \, \frac{31}{12}\log t
\right];
\nonumber\\
\bar{C}_T^{(1)} &=& - \,\frac{C_F}{\pi} \, \frac{637}{144} \, = - \, 1.87744.
\end{eqnarray}
Note that the correction to the coefficient function is very large:
for $\alpha(m_b)=0.22$ it amounts to $-\,41.3\%$.

We now expand the resummed result in powers of $\alpha$ and
compare with the fixed order result,
which is known to full order $\alpha$ \cite{mxspectrum,ndf}:
\begin{equation}
R_T(t;\alpha)\,=\, 1 \,+\, \alpha \, R_T^{(1)}(t) \, + \, \alpha^2 \, R_T^{(2)}(t)
\, + \, O(\alpha^3),
\end{equation}
with
\begin{equation}
R_T^{(1)}(t) \,=\,
- \, \frac{C_F}{\pi}
\left[
\, \frac{1}{2}\log^2 t
\, + \, \frac{31}{12}\log t
\, + \, \frac{637}{144}
\, - \, \frac{97}{18} t \, + \, \frac{25}{18}t^3 \, - \, \frac{61}{144} t^4
\, + \, \left(  \frac{5}{3} t \, - \, \frac{3}{2} t^2 \, + \, \frac{1}{6} t^4 \right)\log t
\right].
\end{equation}
We obtain for the remainder function in first order:
\begin{equation}
\bar{D}_T^{(1)}(t) \, = \,
\frac{97}{18} t - \frac{25}{18}t^3 + \frac{61}{144} t^4
- \left(  \frac{5}{3} t - \frac{3}{2} t^2 + \frac{1}{6} t^4 \right)\log t.
\end{equation}
Since
\begin{equation}
\label{normSigbar}
\bar{\Sigma}_T(1;\,\alpha)\, = \, 1,
\end{equation}
taking $t=1$ in eq.~(\ref{R_T}), we obtain a relation between the coefficient function
and the remainder function in the endpoint:
\begin{equation}
\bar{C}_T(\alpha) \,  = \, 1 \, - \, \bar{D}_T(1;\,\alpha).
\end{equation}
It is a trivial matter to verify that the above relation holds true for our first-order
expressions.

A compact definition of the minimal scheme can be given as follows.
Since
\begin{equation}
\int_0^1
dw \, C_H(w;\alpha) \, \Sigma\left[t/w^2;\,\alpha(w\,m_b)\right] \, = \,
\bar{C}_T(\alpha) \, \bar{\Sigma}_T\left(t;\,\alpha\right),
\end{equation}
the coefficient function in the minimal scheme can be defined taking
$t=1$ in the above equation and using (\ref{normSigbar}):
\begin{equation}
\bar{C}_T(\alpha) \, \equiv \,
\int_0^1 dw \, C_H(w;\alpha) \, \Sigma\left[1/w^2;\,\alpha(w\,m_b)\right].
\end{equation}
The effective form factor then reads:
\begin{equation}
\bar{\Sigma}_T\left(t;\,\alpha\right) \, \equiv \,
\frac{ \int_0^1 dw \, C_H(w;\alpha) \, \Sigma\left[t/w^2;\,\alpha(w\,m_b)\right] }
{\int_0^1 dw \, C_H(w;\alpha) \, \Sigma\left[1/w^2;\,\alpha(w\,m_b)\right]}.
\end{equation}
Let us comment the above result. The effective form factor
involves a convolution over the hadron energy $w$ of the
coefficient function and the universal form factor, which cannot
be reduced to an ordinary product by the standard moment
transform. That is because the variable $w$ enters not only in the
first argument of
$\Sigma=\Sigma\left[t/w^2;\,\alpha(w\,m_b)\right]$ but also in the
scale of the coupling $\alpha=\alpha(w\,m_b)$. Analogously to the
$u$ distribution, there are long-distance effects in the effective
form factor related to small hadron energies $w \ll 1$, which are
suppressed by the coefficient function. In the present case,
however, there is an additional mechanism suppressing the small
energy contributions: since $\Sigma$ is evaluated in $t/w^2$,
small $w$'s correspond to a large argument $u=t/w^2$ of
$\Sigma(u)$, where there are no large logarithms, $\log 1/u \sim
O(1)$, and one is inclusive at the parton level. We may say that
this spectrum is ``double protected'' from the non-perturbative
long-distance effects related to small hadron energies.

The systematic expansion of the form factor is easily obtained
by writing as usual:
\begin{equation}
\bar{\Sigma}_T \, = \, e^{\bar{G}_T},
\end{equation}
one obtains:
\begin{equation}
\bar{G}_T(t;\alpha) \,=\, \sum_{n=1}^{\infty}\sum_{k=1}^{n+1}
\bar{G}_{T n k}\, \alpha^n \, L_t^k,
\end{equation}
where
\begin{equation}
L_t \, \equiv \, \log\frac{1}{t}
\end{equation}
and
\begin{eqnarray}
\bar{G}_{T 12} &=& G_{12};
\\
\bar{G}_{T 11} &=& G_{11} + \frac{5}{6} A_1;
\\
\bar{G}_{T 23} &=& G_{23};
\\
\bar{G}_{T 22} &=& G_{22} + \frac{7}{24} A_1^2 + \frac{5}{6} A_1\beta_0;
\\
\bar{G}_{T 21} &=& G_{21} + \frac{5}{6} A_2
+  A_1^2 \left[ \frac{5}{6} z(2) - \frac{47}{54} \right]
- \frac{23}{36} A_1\beta_0 + \frac{7}{12} A_1 (B_1 + D_1) +
\nonumber\\
&+& \frac{5}{6} \beta_0 D_1
+ A_1 \frac{C_F}{\pi} \left[ \frac{547}{108} - 4 z(3) \right] ;
\\
\bar{G}_{T 34} &=& G_{34};
\\
\bar{G}_{T 33} &=& G_{33} + \frac{83}{648} A_1^3 + \frac{7}{12} A_1^2\beta_0
                     + \frac{10}{9}A_1\beta_0^2;
\\
\bar{G}_{T 32} &=& G_{32} +
\frac{5}{3} \,A_2\,
     \beta_0   +
  \frac{5}{3} \,{\beta_0}^2\,
     D_1  +
\frac{83}{216} \, {A_1}^2
   \left( B_1 \, + \,D_1 \right)
+ {A_1}^2\, \beta_0\,
      \left(
        \frac{35\,z(2)}{12} - \frac{47}
           {27}   \right)
\nonumber\\
&+&  {A_1}^2\, \frac{C_F }{\pi }\,
        \left( \frac{22747}{1728} -
          12\,z(4) \right) +
\frac{7}{12} \, A_1 \, A_2
- \frac{1}{18}
A_1\, \Big( 23\,{\beta_0}^2 - 15 \,\beta_1 \Big) +
\frac{7}{12} A_1\,\beta_0\,
      \Big( B_1 + 2 D_1 \Big) +
\nonumber\\
&+&
        A_1\,\beta_0\,\frac{C_F }{\pi }\,
           \left( \frac{547}{108} -
             4\,z(3)
             \right)
+ {A_1}^3\,
   \left( - \frac{1117}
        {1296}  +
     \frac{7\,z(2)}{12} -
     \frac{5\,z(3)}
      {6} \right).
\end{eqnarray}
For comparison with future higher-order computations, let us
give the explicit values of the coefficients
$\bar{G}_{T ij} \ne G_{ij}$:
\begin{eqnarray}
\bar{G}_{T 11} &=& \frac{31\,C_F}{12\,\pi };
\\
\bar{G}_{T 22} &=& \frac{C_F }{\pi^2}
\, \Bigg[ - \frac{11\,n_f}
       {48} + C_F\,
       \left( \frac{7}{24} -
         \frac{z(2)}{2} \right)
       + C_A\,
       \left( \frac{35}{32} +
         \frac{z(2)}{4} \right)
      \Bigg];
\\
\bar{G}_{T 21} &=&
\frac{C_F}{\pi^2}
\Bigg[ n_f
       \left( \frac{z(2)}{6}- \frac{83}
            {144} \right)
       + C_F\,
       \left( \frac{941}{288} +
         \frac{11}{6} z(2) -
         \frac{7}{2} z(3) \right)  +
      C_A\,
       \left( \frac{107}{32} -
         \frac{11}{6} z(2) -
         \frac{z(3)}{4}
         \right)  \Bigg];~~~~
\\
\bar{G}_{T 33} &=&
\frac{C_F}{\pi^3}
\, \Bigg[ \frac{37}{1296}\,n_f^2 +
    C_F\,C_A\left(\frac{77}{144} - \frac{11\,z(2)}{8}\right) +
        C_F\,n_f\left(-\frac{5}{144} + \frac{z(2)}{4}\right) +\nonumber\\
&&~~~~~  + \, C_A\,n_f \left(-\frac{25}{162} - \frac{z(2)}{12}\right) +
   C_A^2\left( \frac{1057}{5184} + \frac{11\,z(2)}{24}\right) +
    C_F^2\left(\frac{83}{648} + \frac{z(3)}{3} \right)
  \Bigg];
\\
\bar{G}_{T 32} &=&
\frac{C_F}{\pi^3}
\, \Bigg[ \left( \frac{263}{2592} - \frac{z(2)}{36} \right) \,n_f^2 +
    C_F\,C_A\left(\frac{28493}{10368} + \frac{457\,z(2)}{96}
    -\frac{77\,z(3)}{12}+ \frac{5\,z(4)}{4}\right) +
\nonumber\\
&&~~ + \, C_F\,n_f\left(-\frac{2353}{5184} - \frac{47\,z(2)}{48}
     + \frac{11\,z(3)}{12}    \right) +
        C_F^2 \left(\frac{60287}{5184} + \frac{7\,z(2)}{12}
    - \frac{31\,z(3)}{12}  - \frac{47\,z(4)}{4}    \right) +
\nonumber\\
&&~~ + \, C_A\,n_f \left(-\frac{6515}{5184} + \frac{17 z(2)}{36}
    - \frac{z(3)}{24}  \right) +
        C_A^2 \left(\frac{31841}{10368} - \frac{229\,z(2)}{144}
    + \frac{77\,z(3)}{48}  - \frac{11\,z(4)}{16}    \right)
  \Bigg].~~~~~
\end{eqnarray}
Let us make a few remarks:
\begin{itemize}
\item
The coefficient of the single logarithm at $O(\alpha)$ is different
from the previous case as well as from the radiative decay:
\begin{equation}
\bar{G}_{T 11} \, \ne \, G_{U 11} \, = \, G_{11},
\end{equation}
because $\bar{G}_{T 11}$ takes a kinematic contribution from $A_1$, i.e.
from the double logarithm at one loop in $\Sigma$.

The logarithmic structure of $\bar{G}_T$ radically differs from that
of $G$ because the integration variable $w$ enters not only the argument
of the running coupling $\alpha=\alpha(w\,m_b)$
but also the argument of the logarithm $L=\log(t/w^2)$;
\item
If the hard scale was set by the heavy flavor mass, $Q\,=\,m_b$
(a kind of frozen coupling case with respect to the real case)
we would have the following values for the coefficients:
$\bar{G}_{T 22}^{fr} = G_{22} + 7/24 \; A_1^2 + 5/4\; A_1\beta_0$ and
$\bar{G}_{T 33}^{fr} = G_{33} + 83/648 \;A_1^3 + 7/8\; A_1^2\beta_0 +
 35/18\; A_1\beta_0^2$.
\end{itemize}
As discussed in the previous section, in phenomenological studies
one may want to replace the perturbative expression of
$\Sigma(u;\,\alpha)$ with a fit to some experimental data or with
a non-perturbative model. The minimal scheme cannot be used
directly in these circumstances because the effective form factor
$\bar{\Sigma}(t;\,\alpha)$ involves the integration of
$\Sigma(u;\,\alpha)$ in the unphysical region $u>1$. There are
various ways to deal with this problem. One way could be for
example replacing $\Sigma(u;\,\alpha)$ with the non-perturbative
quantity $\Sigma_{np}(u;w)$ for $u\le 1$, while still keeping the
perturbative $\Sigma(u;\,\alpha)$ in the unphysical region $u>1$.
The perturbative form factor is indeed an analytic function of
$u$, which can be continued to any value of $u$. To avoid the
inclusion of the perturbative $\Sigma(u;\,\alpha)$ for $u>1$, let
us consider instead a non-minimal scheme with an effective form
factor defined as:
\begin{equation}
\label{defT}
\Sigma_T\left(t;\,\alpha\right) \,=\,
\frac{ \int_0^1 dw \, C_H(w;\,\alpha) \, \tilde{\Sigma}\left[t/w^2;\,\alpha(w\,m_b)\right] }
{ \int_0^1 dw \, C_H(w,\alpha) },
\end{equation}
where the standard form factor $\Sigma(u;\alpha)$ has been
extended to  arguments larger than one, $u>1$, since:
\begin{equation}
\tilde{\Sigma}(u ;\alpha) \, \equiv \, \Bigg\{
\begin{array}{ll}
\Sigma(u ;\alpha)
&  {\rm for~} u \, \le \, 1;
\\
1
& {\rm for~} u \, > \, 1.
\end{array}
\end{equation}
Because of the definition, it holds:
\begin{equation}
\Sigma_T(1;\,\alpha) \,=\, 1.
\end{equation}
Explicitly one has (cfr. the r.h.s. of eq.~(\ref{goodsplit})):
\begin{equation}
\int_0^1 dw \, C_H(w;\,\alpha) \,
\tilde{\Sigma}\left[t/w^2;\,\alpha(w\,m_b)\right]
\,=\,
\int_0^{\sqrt{t}} dw \, C_H(w;\alpha) \, + \,
\int_{\sqrt{t}}^1 dw \, C_H(w;\alpha) \,
\Sigma\left[t/w^2;\,\alpha(w\,m_b)\right].
\end{equation}
The coefficient function is given in the new scheme by:
\begin{equation}
C_T(\alpha) \, = \,  \int_0^1 dw \, C_H(w,\alpha).
\end{equation}
By inserting in eq.~(\ref{defT}) the perturbative expansions
for $C_H$ and for $\Sigma$, we obtain:
\begin{equation}
\Sigma_T^{(1)}(t) \,=\,
\frac{C_F}{\pi}\Big[
- \frac{1}{2}\log^2 t
- \frac{31}{12}\log t
- \frac{151 - 232\,  t^{3/2} + 81 \, t^2 }{72}
\Big]
\end{equation}
and
\begin{equation}
\label{CT1}
C_T^{(1)} \,=\, - \frac{C_F}{\pi} \, \frac{335}{144} \, = \, - \, 0.98735.
\end{equation}
The first of the above equations shows that $\Sigma_T$, unlike $\bar{\Sigma}_T$,
is not defined in a minimal factorization scheme, i.e. it does not contain only
logarithmic terms $\alpha^n\,L_t^k$, but also contributions of a different form.
Note that the coefficient function in (\ref{CT1}) has a much smaller value than
in the minimal scheme, giving a hint of better convergence of the
perturbative series in the modified scheme.

Matching with the first-order spectrum, one obtains for the remainder function
in the modified scheme:
\begin{equation}
D_T^{(1)}(t) \,=\, \frac{C_F}{\pi}
\left[
  \frac{97}{18} t - \frac{29}{9} t^{3/2} + \frac{9}{8}t^2
- \frac{25}{18}t^3 + \frac{61}{144} t^4
-\left( \frac{5}{3} t - \frac{3}{2} t^2 + \frac{1}{6} t^4 \right)\log t
\right].
\end{equation}
The coefficients of the logarithms in the exponent of the form factor
$G_T$ are the same in the minimal scheme and in the modified one:
\begin{equation}
G_{Tij} \, = \, \bar{G}_{T ij}~~~~~~~~{\rm for}~j\ge 1.
\end{equation}
The non logarithmic coefficients in the modified scheme
are given by ($\bar{G}_{T 10}=0$ and $\bar{G}_{T 20}=0$
by definition of minimal scheme):
\begin{eqnarray}
G_{T 10} &=&
-\,\frac{23}{36} \,{A_1} \, + \,
  \frac{5}{6}\,({B_1} + D_1);
\\
G_{T 20} &=&
-\,\frac{23}{36} \,A_2 \, + \,
  \frac{7}{24}\,(B_1 + D_1)^2 +
  \frac{5}{6}\,({B_2}+D_2) +
   \frac{23}{36} \, \beta_0 \, B_1   +
\nonumber\\
&+& {A_1}\,
   \left( {B_1} + D_1 \right)
      \left( -\,\frac{47}{54}   +
        \frac{5\,z(2)}{6}
         \right)  +
  A_1^2\,
   \left( \frac{2057}{2592} -
     \frac{23\,z(2)}{36} +
     \frac{5\,z(3)}{6} \right)  +
\nonumber\\
&+&  \frac{C_F}{\pi}\,
     \left( {B_1}\,+D_1 \right)
        \left( \frac{547}{108} - 4\,z(3)
          \right)  +
      \frac{C_F}{\pi} \, A_1 \,
        \left( -\, \frac{90121}{5184}
              + \frac{10\,z(3)}{3} +
          12\,z(4) \right).
\end{eqnarray}
Explicitly:
\begin{eqnarray}
G_{T 10} &=& - \, \frac{C_F }{\pi} \, \frac{151}{72};
\\
G_{T 20} &=&
\frac{C_F}{\pi^2}
\Bigg\{
C_A\, \left( \, - \frac{344}
        {81}  + \frac{3}{2}\,z(2) +
     \frac{5}{24}\,z(3) \right)  +
  n_f\,\left( \frac{971}{1296} -
     \frac{5}{36}\,z(2)
     \right)  \, +
\nonumber\\
&&~~~~ + \, C_F\,
   \left( - \frac{79889}{3456}  -
     \frac{53}{36}\,z(2) +
     \frac{119}{12}\,z(3) + 12\,z(4) \right)
\Bigg\}.
\end{eqnarray}
The relations between the coefficient function in the minimal scheme
and in the modified one are obtained imposing that
\begin{equation}
C_T \, e^{G_T} \, = \, \bar{C}_T \, e^{\bar{G}_T} \, + \, O(t;\alpha)
\end{equation}
and read:
\begin{eqnarray}
C_T^{(1)} &=& \bar{C}_T^{(1)} \, - \, G_{T10};
\nonumber\\
 C_T^{(2)} &=& \bar{C}_T^{(2)} \, - \,
\bar{C}_T^{(1)} \, G_{T10}
\, + \, \frac{1}{2}G_{T10}^2 \, - \, G_{T20}.
\end{eqnarray}
The first equation can be directly verified by inserting our
first-order expressions.
The second order correction to the coefficient function
is unknown at present in either scheme and its determination
requires a full two-loop calculation; the second of the above
equations simply allows us to transform the coefficient function from
one scheme to another one.

The Babar collaboration has recently presented the differential spectrum in
$t$ \cite{babarmx}, which is obtained from the previous one by differentiation:
\begin{equation}
\frac{1}{\Gamma}\frac{d\Gamma}{dt} \,=\, \frac{d}{dt} R_T(t).
\end{equation}
Due to their relevance, let us present explicit formulas. The
resummed spectrum reads:
\begin{equation}
\frac{1}{\Gamma}\frac{d\Gamma}{dt} \, = \, C_T(\alpha) \,
\sigma_T(t;\,\alpha) \, + \, d_T(t;\,\alpha)
\end{equation}
where
\begin{itemize}
\item
the coefficient function is the same as in the event fraction,
since it is independent on $t$;
\item
the effective form factor is
\begin{equation}
\sigma_T(t;\,\alpha) \,=\, \frac{d}{dt} \Sigma_T(t;\,\alpha).
\end{equation}
More explicitly:
\begin{eqnarray}
\label{sigTdef}
\sigma_T(t;\,\alpha)
&=& \frac{\int_{\sqrt{t}}^1 d w/w^2 \, C_H(w;\,\alpha) \, \sigma[t/w^2;\,\alpha(w\,m_b)]}
{\int_0^1 d w \, C_H(w;\,\alpha)}
\nonumber\\
&=& \frac{\int_t^1 du/(2\sqrt{t \, u})\,
C_H\left(\sqrt{t/u};\,\alpha(w \, m_b)\right) \,
\sigma\left[u;\,\alpha\left(m_b\sqrt{t/u} \right)\right]}
{\int_0^1 d w \, C_H(w;\,\alpha)},
\end{eqnarray}
where a double representation as an integral over $w$ or over $u$
respectively has been given;
\item
the remainder function is:
\begin{equation}
d_T(t;\,\alpha) \,=\, \frac{d}{dt} D_T(t;\,\alpha)
\,=\, \alpha \, d_T^{(1)}(t) \, + \, \alpha^2 \, d_T^{(2)}(t) \, +\, O(\alpha^3).
\end{equation}
The explicit value of the first-order correction is:
\begin{equation}
d_T^{(1)}(t) \, = \, \frac{C_F}{\pi} \left[
\frac{67}{18} - \frac{29}{6}\sqrt{t} + \frac{15}{4} t -  \frac{25}{6} t^2
+ \frac{55}{36} t^3 -\frac{5}{3} \log t + 3t\log t - \frac{2}{3} t^3 \log t
\right].
\end{equation}
\end{itemize}
Let us make a comment about the non-perturbative effects entering
the differential mass distribution. The expression for the
effective form factor $\sigma_T(t;\,\alpha)$ at the second member
of eq.~(\ref{sigTdef}) involves an integration over $w$ from
$\sqrt{t}$ up to one. Since the running coupling is evaluated in
$Q=w\,m_b$, the smallest hard scale contributing to the
distribution is
\begin{equation}
Q_{min} \, = \, \sqrt{t} \,\, m_b \, = \, m_X.
\end{equation}
In order to avoid the infrared pole in the coupling --- the well-known Landau pole
--- implying a breakdown of the perturbative scheme, one has to impose the condition
\begin{equation}
 m_X \, \gg \, \Lambda_{QCD},
\end{equation}
which is also very reasonable from the physical viewpoint.
Resummed perturbation theory therefore signals that the hadron
mass distribution cannot be computed for hadron masses of the
order of the hadron scale because of the appearance  of the Landau
pole\footnote{The universal QCD form factor
$\sigma\left[u;\,\alpha(Q)\right]$ has an infrared singularity at
$ u = \exp \left[-1/(2 \, \beta_0 \, \alpha(Q)) \right]$, related
to the Landau pole for $m_X\approx\sqrt{\Lambda_{QCD}Q}\gg
\Lambda_{QCD}$. We assume that the latter has been regulated in
some way; for a recent discussion see for example
\cite{nostrareg}.}.

\section{Conclusions}
\label{concl}

In this work we have presented next-to-leading resummed expressions
for the distribution in the final hadron mass/energy ratio
and for the distribution in the invariant hadron mass
in the semileptonic decays
\begin{equation}
\label{prima}
B \, \rightarrow \, X_u \, + \, l \, + \, \nu.
\end{equation}
By expanding our formulas, we have obtained the coefficients of
all the infrared logarithms to $O(\alpha^2)$ and of the leading
ones to $O(\alpha^3)$.
These two spectra have different logarithmic structures
from each other, which are both different also from that one in the
radiative decay (\ref{raddec}).
That occurs because these spectra involve integration over the
total hadron energy $E_X$, which sets the hard scale $Q$ of the
hadronic subprocess in (\ref{prima}):
\begin{equation}
Q \,  = \, 2 E_X.
\end{equation}
Long distance effects manifest in perturbation theory in the form of large
infrared logarithms, which are usually factorized and resummed in
QCD form factors.
Universality of long-distance effects therefore shows up
in perturbation theory as the occurrence of equal form factors
in different distributions.
That implies that the spectra we have considered in this work
have different long-distance effects from each other as well as from the
radiative decay (\ref{raddec}).
There is no simple connection between these semileptonic spectra
and the hadron mass distribution in radiative decays.
For both spectra, we have introduced effective, i.e. process
dependent, form factors, which factorize the large
logarithms to all orders in perturbation theory.
These effective form factors can also be computed in a phenomenological way
by inserting, in place of the perturbative QCD form factor
$\Sigma\left[u;\,\alpha(w\,m_b)\right]$,
the form factor $\Sigma_{np}(u;\,w)$ computed with a non-perturbative model
or a fit to experimental data.
In the case of the distribution in the hadron mass squared $t$, we have also
considered a non-minimal factorization-resummation scheme, which seems
to have better convergence properties of the perturbative series with respect
to the minimal one.

There are other important semileptonic spectra which have similar properties
to those of the distributions considered here, i.e. long distance effects
which cannot be factorized into a process independent form factor, to be extracted
for example from the radiative decay \cite{inpreparation}.

Finally, we have also shown that the cancellation of long-distance effects
in the ratio constructed in \cite{ucg} occurs only in leading order
while it is violated at the level of $\alpha^2\log^2\left(\frac{1}{u}\right)$ terms.

\end{document}